\documentclass[conference]{IEEEtran}
\IEEEoverridecommandlockouts

\usepackage[latin9]{inputenc}
\usepackage{amsmath,graphicx}
\usepackage{amssymb,amsmath,mathtools}
\usepackage{subfigure}
\usepackage{graphicx,epsfig}
\usepackage{empheq}
\usepackage{cite}
\usepackage{color}
\usepackage[usestackEOL]{stackengine}
\usepackage[normalem]{ulem}
\usepackage{balance}

\usepackage{physics}
\usepackage{amsmath}
\usepackage{tikz}
\usepackage{mathdots}
\usepackage{yhmath}
\usepackage{cancel}
\usepackage{color}
\usepackage{siunitx}
\usepackage{array}
\usepackage{multirow}
\usepackage{amssymb}
\usepackage{gensymb}
\usepackage{tabularx}
\usepackage{extarrows}
\usepackage{booktabs}
\usetikzlibrary{fadings}
\usetikzlibrary{patterns}
\usetikzlibrary{shadows.blur}
\usetikzlibrary{shapes}

\usepackage{xcolor}
\usepackage{algpseudocode}
\usepackage{algorithm}
\usepackage{bm}
\usepackage{tikz}
\usetikzlibrary{decorations.pathreplacing,calc}
\newcommand{\tikzmark}[1]{\tikz[overlay,remember picture] \node (#1) {};}

\makeatletter
    \algrenewcommand\alglinenumber[1]{\tikzmark{\arabic{ALG@line}}\tiny#1:}
\makeatother

\makeatletter

\newcommand{\Rmnum}[1]{\expandafter\@slowromancap\romannumeral #1@}

\makeatletter

\let \bs = \boldsymbol

\def \p1#1{#1^{-1}}

\def \~#1{\tilde{#1}}

\newcommand{\mt}[1]{}
\let \mt=\mathrm

\newcommand\dss{\bgroup\markoverwith{\textcolor{blue}{\rule[0.5ex]{2pt}{0.4pt}}}\ULon}

\ifCLASSINFOpdf
\else
\fi
\begin{document}
%
\title{Space Observation by the Australia Telescope Compact Array: Performance Characterization using GPS Satellite Observation}
\author{\IEEEauthorblockN{Hamed Nosrati\IEEEauthorrefmark{1}, Stephanie Smith, Douglas B. Hayman}
\IEEEauthorblockA{ CSIRO Space \& Astronomy, Australia\\ \{Hamed.Nosrati,Stephanie.Smith,Douglas.Hayman\}@csiro.au\\\IEEEauthorrefmark{1} Corresponding author
}
}


\maketitle

\begin{abstract}
In order to operationalize the Australia Telescope Compact Array (ATCA) for space situational awareness (SSA) applications, we develop a system model for range and direction of arrival (DOA) estimation based on the interferometric data. We employ the observational data collected from global positioning system (GPS) satellites to evaluate the developed model and demonstrate that, compared to a priori location propagated from the most recent two-line element (TLE), both range and direction information are improved significantly.    
\end{abstract}
\begin{IEEEkeywords}
Space Situational Awareness, Direction of Arrival Estimation, Radio Interferometry, Space Debris
\end{IEEEkeywords}

\IEEEpeerreviewmaketitle

\section{Introduction}\label{intro}

The increasing congestion of resident space objects (RSOs) demands more sensors to provide observational data for cataloging in the context of Space Situational Awareness (SSA) and Space Traffic Management (STM)\cite{Maclay_2021}\cite{Muelhaupt2019}. In addition to the expanding network of dedicated optical and radio frequency (RF) sensors, radio telescopes, which are mainly being used for scientific astronomical observation, offer a largely untapped opportunity for delivering SSA data products. A radio telescope operationalized for SSA application could provide various measurements, i.e., location, range, and range rate, individually or coordinated with a transmitter.

The Australia Telescope Compact Array (ATCA) is an array of six twenty-two-meter Cassegrain reflector antennas located at the Paul Wild Observatory near Narrabri in New South Wales (NSW), Australia. ATCA is primarily used for astronomical observations as a stand-alone interferometer or as part of a larger VLBI network with existing receivers covering bands within 1-115~GHz. To study the performance of the ATCA correlator array in interferometry mode for space observations and establish an operational SSA mode, a system model and a sensitivity model are required. 
In this work, we consider the interferometric observation of medium earth orbit (MEO) objects and describe a system model for observations required for orbit determination. In particular, we develop a signal and system model for estimating the range and direction of arrival (DOA) based on the multiple signal classification (MUSIC) parameter estimation technique. We use the correlator's output and adopt interferometry equations to refocus the interferometer on a grid of ranges near the location of interest to establish the measurement vector containing the unknown parameters.   



Interferometric observations of global positioning system (GPS) satellites are used to evaluate the system model proposed for localization by comparing the location estimates with those of the \emph{ground truth}; the precise GPS location data points published by the US National Geospatial-Intelligence Agency (NGA) \cite{nga}. In each observation, we use the most recent two-line element (TLE) published by North American Aerospace Defence Command (NORAD) as the cue. The beamwidth of our dishes, about 0.6 degrees, comprises our \emph{search area} centered on these a priori coordinates. We then refine the trajectory parameters based on our observations and compare them to the precise ephemerides.


\section{System Description}

As described above, we use ATCA for observations of objects in the earth's orbit. Among ATCA's various operational modes, the interferometry and tied array ones are the most frequently used. In the interferometry mode, the correlation of each pair of antennas is computed per integration cycle per frequency subband. However, in the tied array mode, the \emph{raw voltages} of a set of elements are combined while tracking delays for a particular phase center from a single terrestrial coordinate. Being limited to a single terrestrial coordinate means that the set of tied elements provides a single array as big as the corresponding baseline, which makes it a good candidate for either Doppler processing as an individual sensor or interferometry with other telescopes in a very long baseline interferometry setup. Since we are interested in using ATCA on its own for localization and orbit determination, we utilize the interferometry mode over a narrow band and a short integration cycle.  




 \begin{figure*}[!t]
    \centering
    \includegraphics[trim=0 100 0 0,clip,width=11cm]{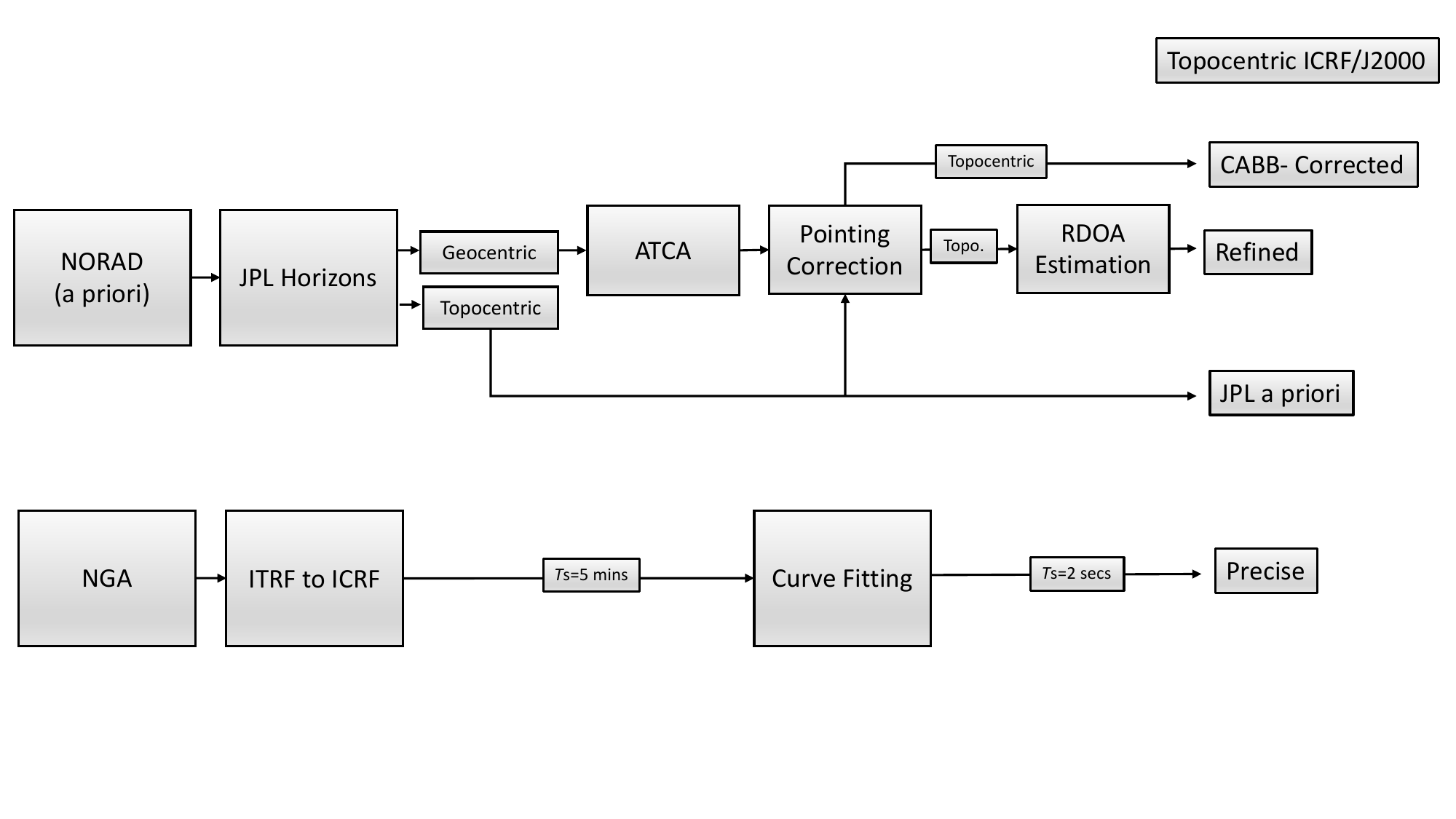}
    \caption{System model and architecture of the evaluation method. }
    \label{fig:ev_arch}
\end{figure*}

\section{Signal Model}\label{sec:sigmod}
In this section, we develop a signal model for range and DOA estimation based on interferometric data. The output of the correlator is the visibilities, i.e., correlated voltages for different antenna pairs for a tracked phase center on the celestial sphere.  For a set of baselines defined in the $uvw$ coordinate system within the terrestrial frame, the interferometer's output at a particular subband for a source located in the far-field region is modeled as
\begin{align}\label{eq:vis_uv}
    \nu( u,v)&=\int\limits_{-\infty}^{\infty} \int\limits_{-\infty}^{\infty}A_{f}(l,m)I_f(l,m) e^{-j2\pi(ul+vm))}\,dl\,dm,
\end{align}
where $l,m$ denote the direction cosines of a vector in $uv$ coordinates. $A_{f}(l,m)$ is the effective collecting area of each antenna, and $I_{f}(l,m)$ denotes the intensity distribution at frequency $f$ that is the center frequency of the underlying subband. This model is employed in radio astronomy for emissions from locations on the celestial sphere,  and the sky brightness, i.e., $I_{f}(l,m)$ is measured by taking an inverse Fourier transform of (\ref{eq:vis_uv}). For space observations, the emissions are located much closer to the array; hence the visibilities need to be refocused on a particular distance and tailored for the near-field region transmission. Motivated by \cite{cornwell2004correction}, \cite{lazio2009near}, and \cite{carter1989refocusing}, for visibilities refocused on distance $d$ we develop and employ the following model

%

\begin{align} \label{eq:int_steer}
\nonumber\nu_{ij}( u,v,d)=\int\limits_{-\infty}^{\infty}\int\limits_{-\infty}^{\infty}A_{f}(l,m,d)I_f(&l,m,d)c_{i}(l,m,d)\times\\
&c_{j}^H(l,m,d)\,dl\,dm,
\end{align}
where $d$ is the distance from the reference $uv$ origin, $c$ is a vector, defined to decompose the visibility as the multiplication of phase components of the received signal at the $i$-th and $j$-th antennas based on the first-order near-field geometric model as follows
\begin{align}\label{eq:c_i}
c_{i}(l,m,d)=\exp\left(-j2\pi\left(-d_i^d+x_i^dl+y_i^dm-z_i\right)\right),
\end{align}
The core idea here is to presume different locations, i.e. ($x,y,z$) for the antennas in $uvw$ coordinates, given different assumptions on the distance. For a given distance $d$, we calculate the corresponding locations $x_i^d,y_i^d$ as
\begin{align}\label{eq:x_i}
x_i^d=\frac{d x_i}{d_i^d}\;\;\;y_i^d=\frac{d y_i}{d_i^d},
\end{align}
with $d_i^d$ being
\begin{align}\label{eq:d_i}
d_i^d=\sqrt{x_i^2+y_i^2+(z_i-d)^2}
\end{align}

To proceed, we discretize (\ref{eq:int_steer}) by assuming that there are $K$ spatially-incoherent sources located at $(l_\ell,m_\ell,d_\ell)\;\mt{s.t.}\;\ell \in K$, contributing to the measured intensity 

\begin{align} 
\nonumber\nu_{ij}(u,v,d)=\sum\limits_{\ell=1}^K A_{f}(l_\ell,m_\ell,d_\ell)&I_f(l_\ell,m_\ell,d_\ell)\\ &c_{i}(l_\ell,m_\ell,d_\ell)c_{j}^H(l_\ell,m_\ell,d_\ell),
\end{align}
and redefine it as a measurement vector 
\begin{align} \label{eq:meas_vect}
  \nu(u,v,d)&=\bs c_{i}\,\bs b\, \bs c^H_{j},
\end{align}
where
\begin{align}\label{eq:avec}
\nonumber\bs c_{i}&=\Big[c_{i}(l_1,m_1,d_1),\hdots,c_{i}(l_K,m_K,d_K)\Big]\;\;\; 1\times K\\
\nonumber\bs b&=\mt{diag}\Big( A_{f}(l_1,m_1,d_1)I_f(l_1,m_1,d_1),\hdots,\\ &\hspace{2cm}A_{f}(l_K,m_K,d_K)I_f(l_K,m_K,d_K)\Big)^T.\;\;\;K\times K
\end{align}
The correlation of signals received by antennas $i$ and $j$ denoted by (\ref{eq:meas_vect}) can be thought of as an entry of a covariance matrix, i.e., $R_k[i,j]=\nu_{ij}(u,v,d)$.
In practice, the measurement vector in (\ref{eq:meas_vect}) is subject to measurement noise and phase uncertainty. For simplicity, we assume that after calibration, the phase uncertainty is negligible. By taking only the noise  into account, we have
\begin{align}
\bs R[i,j]=\bs c_{i}\,\bs b\, \bs c_{j}^H+n_{i,j}
\end{align}
where $n_{i,j}$ is an additive term resulting from the measurement noise in each antenna. If we consider a zero mean and spatially white measurement noise, then we have
\begin{align}
n_{i,j}=\begin{cases}0\;\;\;\;&i\neq j\\
\sigma ^{2}\;\;\;\;&i= j.\end{cases}
\end{align}
Finally, by stacking the visibilities (correlation lags) on top of each other, we can cast the entire measurement at epoch $t$ in a matrix form as
\begin{align}\label{eq:meas_model}
    \nonumber\bs R(t)=\mathcal{A}(\bs l,\bs m,\bs d)\bs b(\bs l,\bs m,\bs d)\mathcal{A}^H(\bs l,\bs m,\bs d)+\sigma\bs I_N
\end{align}
where
\begin{align}
    \nonumber\mathcal A(t)&=[\bs c^T_{1},\bs c^T_{2},\hdots,\bs c^T_{N}]^T,\;N\times K\\
    &=[\bs a_{1};\bs a_{2};\hdots;\bs a_{K}],\;N\times K
\end{align}
Note that the row vector $\bs a_i$ denotes the array steering vector towards the \mbox{$i$-th} intensity source as opposed to $\bs c_i$, which shows the phase vector of all $K$ sources at the \mbox{$i$-th} lag. The steering vector is related to $c$ as follows
\begin{align}
\bs a_i=[c_1(l_i,m_i,d_i),\hdots,c_N(l_i,m_i,d_i)]^T\;N\times 1.    
\end{align}

The covariance matrix in (\ref{eq:meas_model}) can be associated with a linear measurement model in an array of antennas as
\begin{align}\label{eq:sig_meas_model}
    \bs y(t)=\mathcal{A}(\bs l,\bs m,\bs d)\bs b(\bs l,\bs m,\bs d)+\bs n(t)
\end{align}

Given the measurement model  (\ref{eq:sig_meas_model}) for the $K^{\text{th}}$ source and assuming white Gaussian noise, $(\bs l,\bs m,\bs d)$ and $\bs b_k(\bs l,\bs m,\bs d)$  are the unknown parameters to identify. In this work, we focus on the location parameters and assume that $\bs b_k$ can be identified accordingly. Noting that the covariance matrix is provided by the interferometer, subspace methods can be used directly on the covariance matrix to identify the parameters. We employ the MUSIC algorithm to form a 3D spectrum as
 \begin{align}
     I_{M}(l,m,d)=\left(\bs a^T\left(l,m,d\right)\left(\bs R_n^s\right)^{-1}\bs a\left(l,m,d\right)\right)^{-1}
 \end{align}
 where $\bs R_n^s$ denotes the covariance matrix associated with the noise subspace. The noise subspace is given by the eigenvectors that span the noise subspace. It is worth noting that $I_{M}$ is indeed what the radio astronomy community refers to as a \emph{dirty} beam provided by the MUSIC estimator, and the $K$ peaks of the MUSIC dirty beam determine the location of $K$ sources, i.e., $(\bs l,\bs m,\bs d)$.

\section{Evaluation Method}
A block diagram of the architecture of evaluation is presented in Fig.~\ref{fig:ev_arch}. In essence, we initiate the observation using the most recent TLE. Then we refine the location through the method described in Section \ref{sec:sigmod} and compare it to the location derived from the precise \emph{ground truth} orbital parameters.
To set up the observation, we supply the most recent TLE published by NORAD to the Jet Propulsion Lab (JPL) Horizons System \cite{horizons} and retrieve the geocentric right ascension (RA) and declination (Dec) of the RSO of interest. Then, ATCA's pointing software converts the geocentric coordinates to topocentric coordinates referenced to the correlator's terrestrial coordinates. ATCA tracks the computed topocentric coordinate, i.e., phase center, through a combination of antenna pointing and delay tracking and provides the visibilities for the observation. 
    
After carrying out the observation, we do a phase center evaluation step in order to make sure that the delays have been tracked correctly. 
This has been found necessary because we have detected a systematic error in the ATCA system. This error only becomes apparent for objects relatively close to the earth, and the cause has yet to be resolved.
We compare the phase center of the observation as reported by ATCA against the a priori topocentric coordinates given by the Horizons System. The phase center systematic error is then accounted for, and the visibilities are rotated appropriately so that the phase center matches that of the a priori pointing data. 


In order to evaluate the accuracy of the estimated location based on our observations, we use the ephemeris published by the The National Geospatial-Intelligence Agency (NGA). These ephemerides are generated after the satellite pass and are produced from GPS data at a number of monitoring stations across the globe  \cite{nga}, and so we use them as our \emph{ground truth}. These precise ephemerides are available in either the satellite's center of mass (COM) or antenna's phase center (APC) positions in five-minute intervals in the ITRF and GPS time. Given that we run a passive RF observation of the GPS signal, we use the APC ITRF positions and tailor the frame and time interval to enable comparing that with ATCA's output. We first convert the APC positions from ITRF to international celestial reference system and frame (ICRF) employing Algorithm (23)\footnote{ Algorithm (23) rotates ITRF to GCRF based on IAU-2000 conventions. Noting that ITRF and WGS-84 agree at the centimeter level and the difference between ICRF and J2000 is at the micro arcsecond level, we assume that the rotation is accurate enough in our context.  } in \cite{valladobook}. Also, to make the ephemerides comparable to that of ATCA, which is observed every 2 seconds at a universal time (UT), we apply the leap second and then interpolate the APC ephemeris to have datapoints in 2-second intervals. 


\section{Results}

\begin{figure}[!t]
        \includegraphics[width=8.5cm]{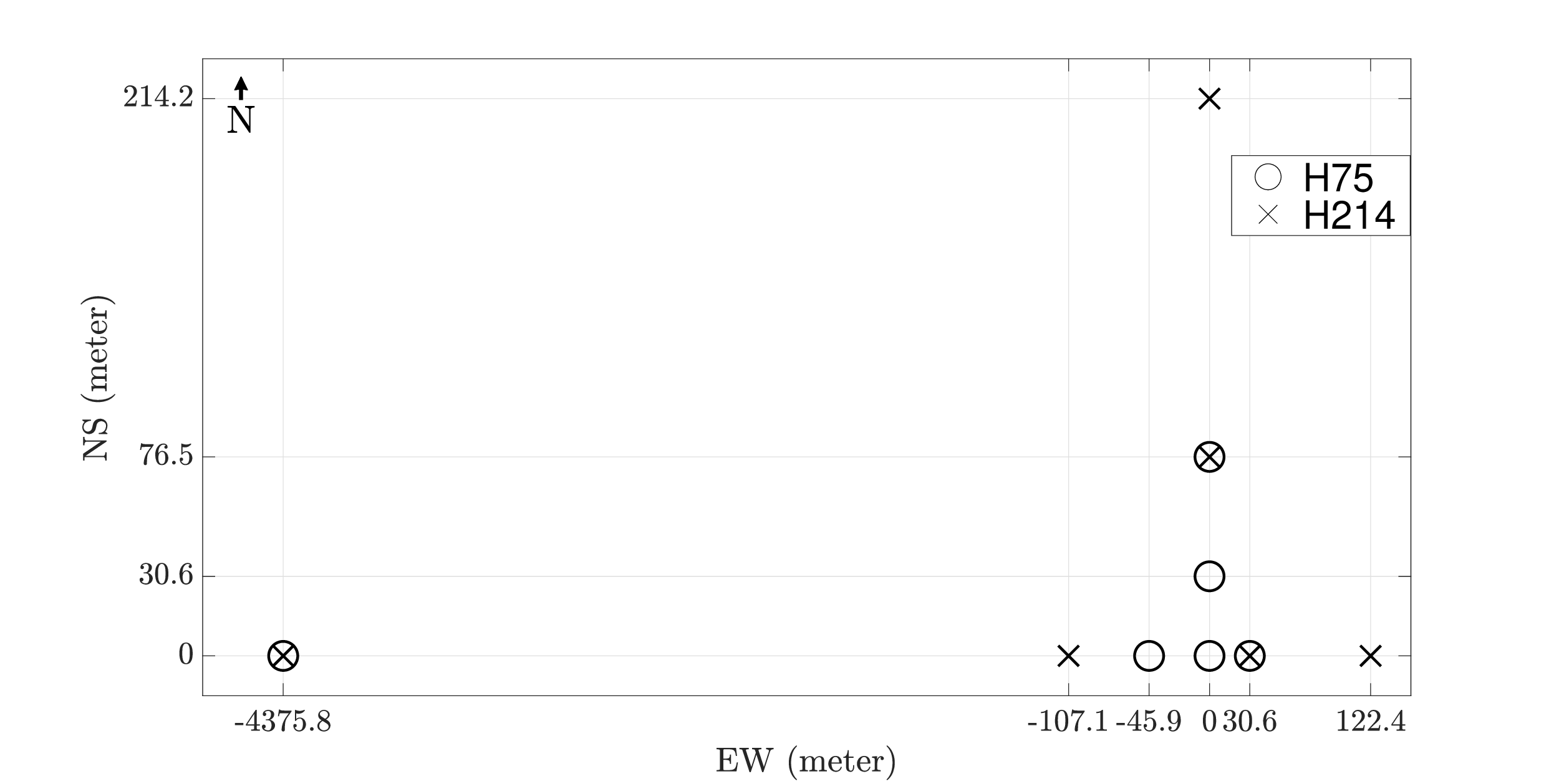}
        \caption{Array configurations during the observations.}
        \label{fig:array_conf}
        \includegraphics[width=8.5cm]{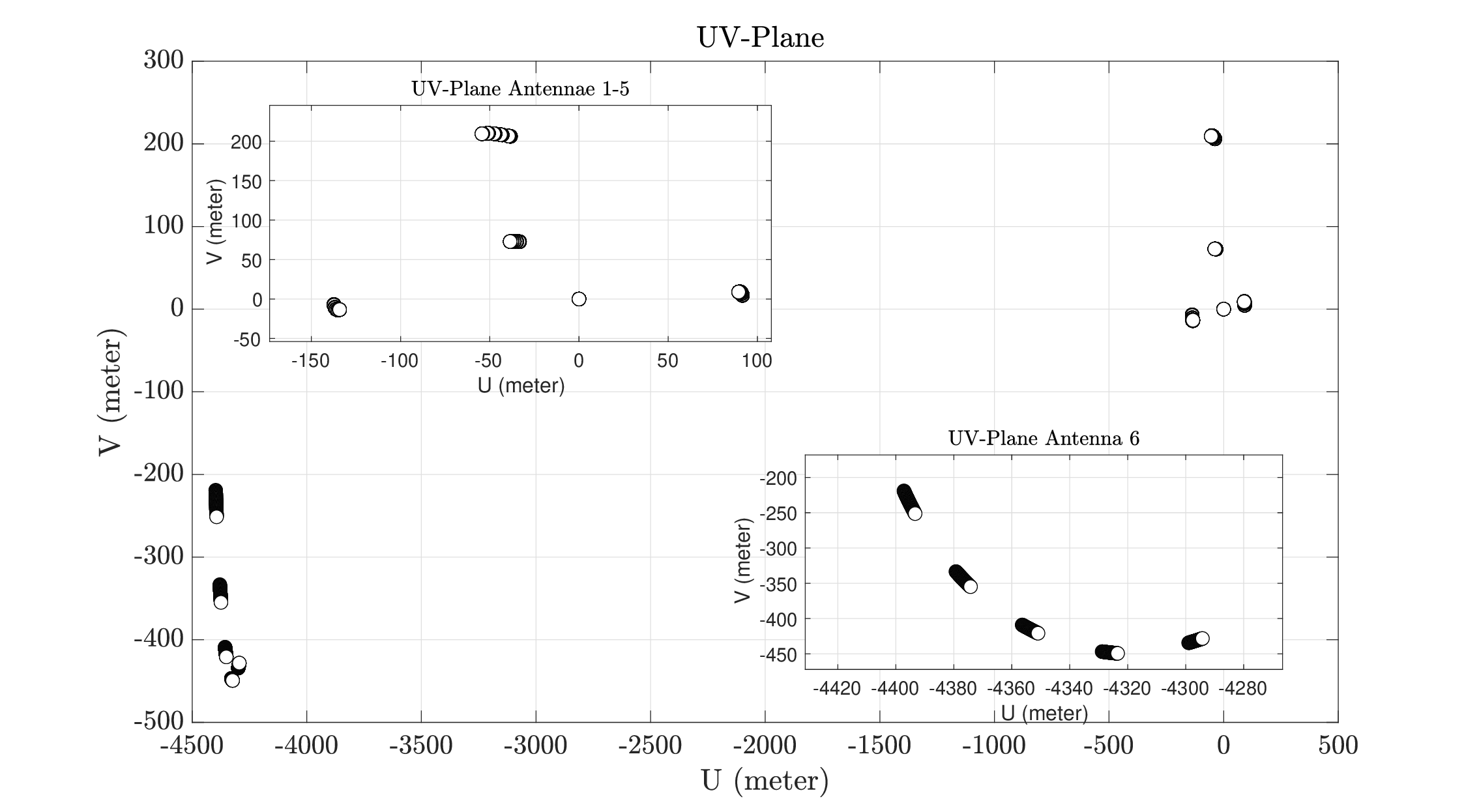}
        \caption{$uv$ coordinates for August observation, configuration H214.}
        \label{fig:uv_coord}
    \end{figure}

We ran observations on 1 June 2021 and 30 August 2021. The ATCA's array and correlator configurations are listed in Table.~\ref{tab:listoparams}. The array configuration are also shown in Fig. \ref{fig:array_conf}.
    \begin{table}[!t]
    \caption{list of ATCA's configuration parameters for the observations.}
    \label{tab:listoparams}
    \centering
    \begin{tabular}{|m{3cm}|m{2cm}|m{2cm}|}
             \hline
        \textbf{Observation} & \textbf{1 Jun 2021}& \textbf{25 Aug 2021} \\
         \hline
 Array Configuration  & H75 & H214 \\
\hline
 Bandwidth  & 488 Hz & 488 Hz  \\
 \hline
 PRN  & 22 & 9  \\

\hline
    \end{tabular}
\end{table}    
    
        \begin{figure*}[!h]
        \centering
        \includegraphics[trim=200 100 0 0,clip,height=21cm]{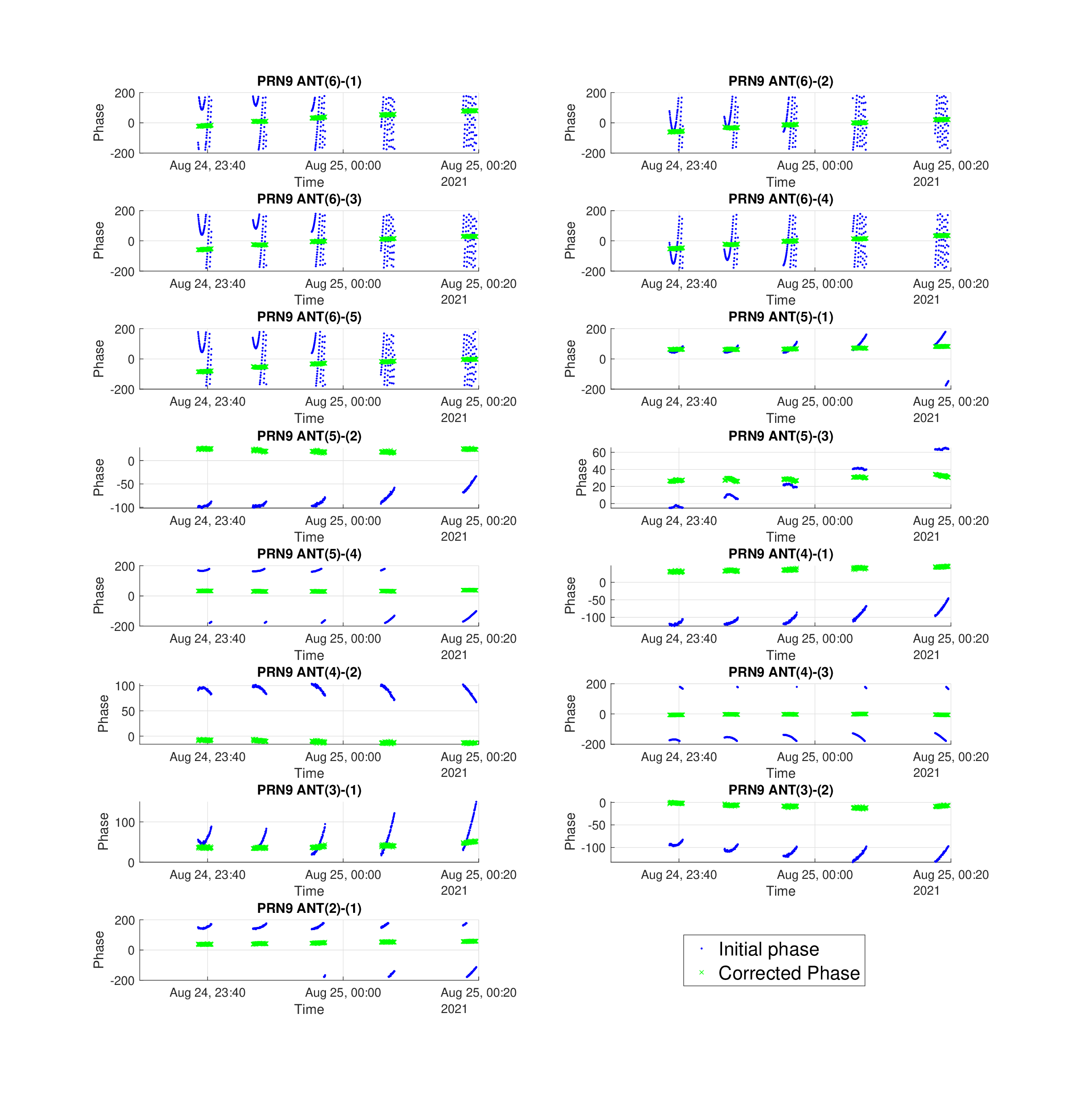}
        \caption{Impact of delay tracking correction in different baselines.}
        \label{fig:corrected_vis}
    \end{figure*} 

The first stage of post-processing is delay tracking correction as we have verified that there is a systemic error in this system while observing orbital objects. To address this, we devise a \emph{re-rotation} procedure to make sure that the phase center has been tracked properly. We first form the correct $uvw$ coordinates based on the TLE data and then compare them with the coordinates that ATCA computes. The $uv$ coordinates computed for the August observation are shown in Fig.~\ref{fig:uv_coord} as an example. Given that the delay tracking depends on the $w$ coordinates, we calculate the difference of the $w$ parameter between correct and assumed baselines and then add the residual phase to each visibility pair. This process is demonstrated in Fig.~\ref{fig:corrected_vis} where there are slopes in the initial phase of the visibility pairs, and it even wraps in longer baselines. After correction, the phases become relatively flat and stable.           

    \begin{figure}[!h]
        \includegraphics[width=8.2cm]{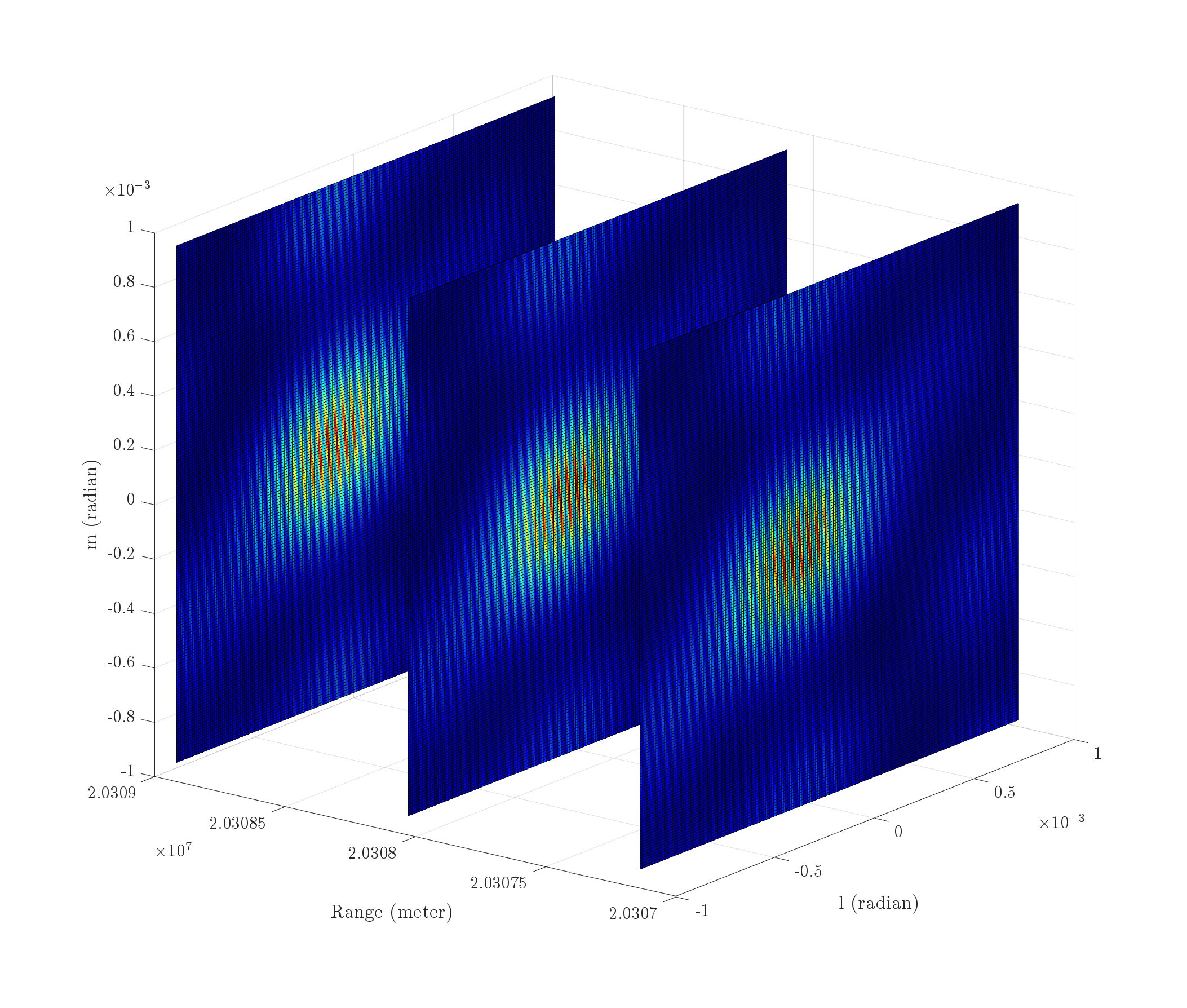}
        \caption{Different slices of the 3D direction-range spectrum.}
        \label{fig:range_bins}
    \end{figure}  
    
        \begin{figure}[!h]
        \includegraphics[width=8.2cm]{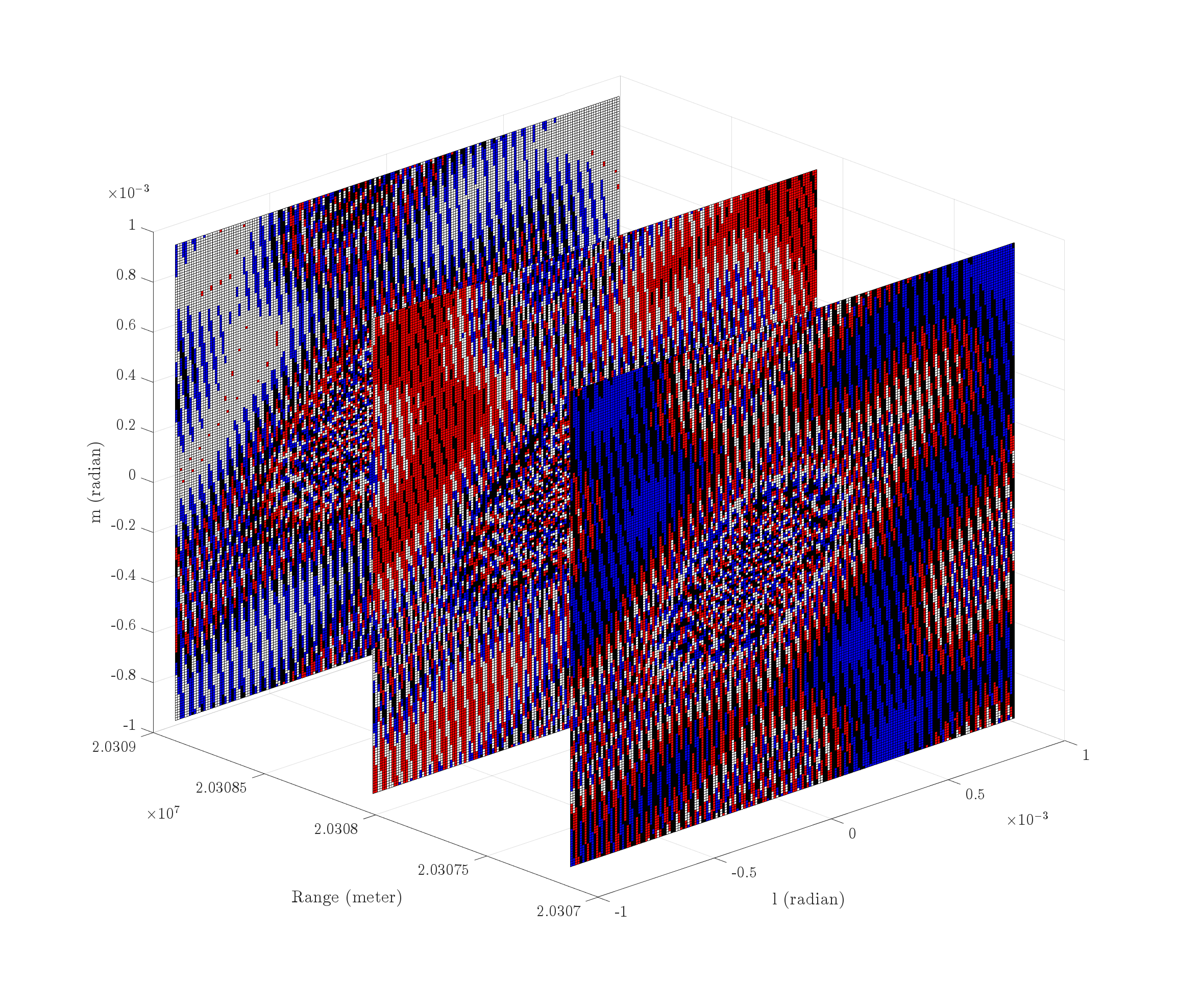}
        \caption{Different slices of the normalized 3D direction-range spectrum.}
        \label{fig:range_bins_diff}
    \end{figure}   
 
After correcting the delay tracking, we plug the visibilities into the range and DOA estimator integration cycle by integration cycle to estimate the range and DOA. As mentioned previously, we refocus the array on the near-field region, which allows us to estimate the range as well as direction. The output of the estimator is a 3D matrix including estimated power at $l,m$ directions as well as $d$ ranges. An example of this matrix is presented in Fig.~\ref{fig:range_bins} where three slices at three range values are shown. In order to highlight the difference in direction spectrum at different ranges, we show a normalized view of the 3D power spectrum in Fig.~\ref{fig:range_bins_diff} where the average direction spectrum has been subtracted from each slice. 

        \begin{figure}[!h]
        \centering
        \includegraphics[width=8.5cm]{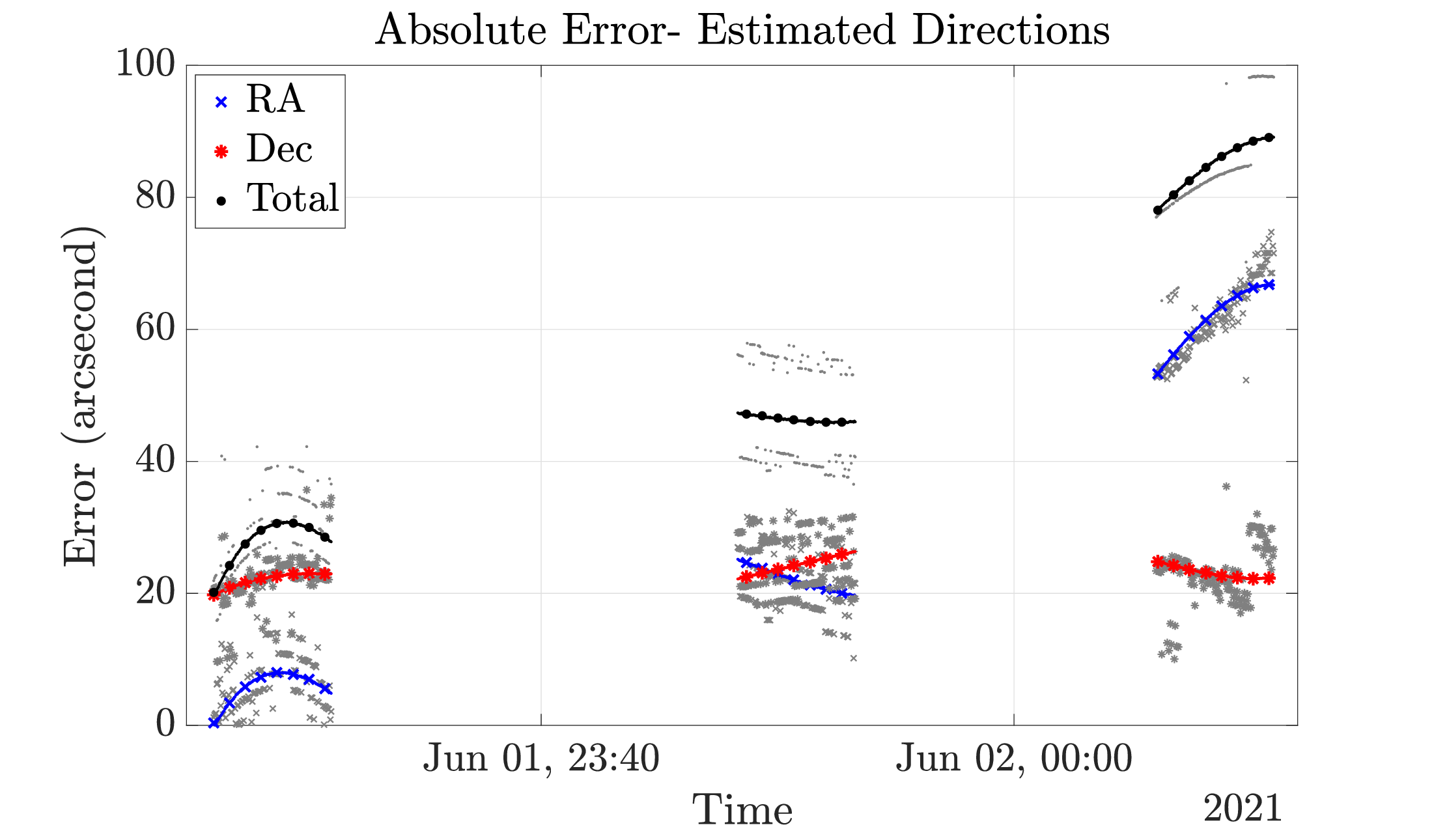}
        \caption{Absolute error of the estimated RA and Dec for the June observation.}
        \label{fig:error_estimated_dir_jun}
\centering
        \includegraphics[width=8.5cm]{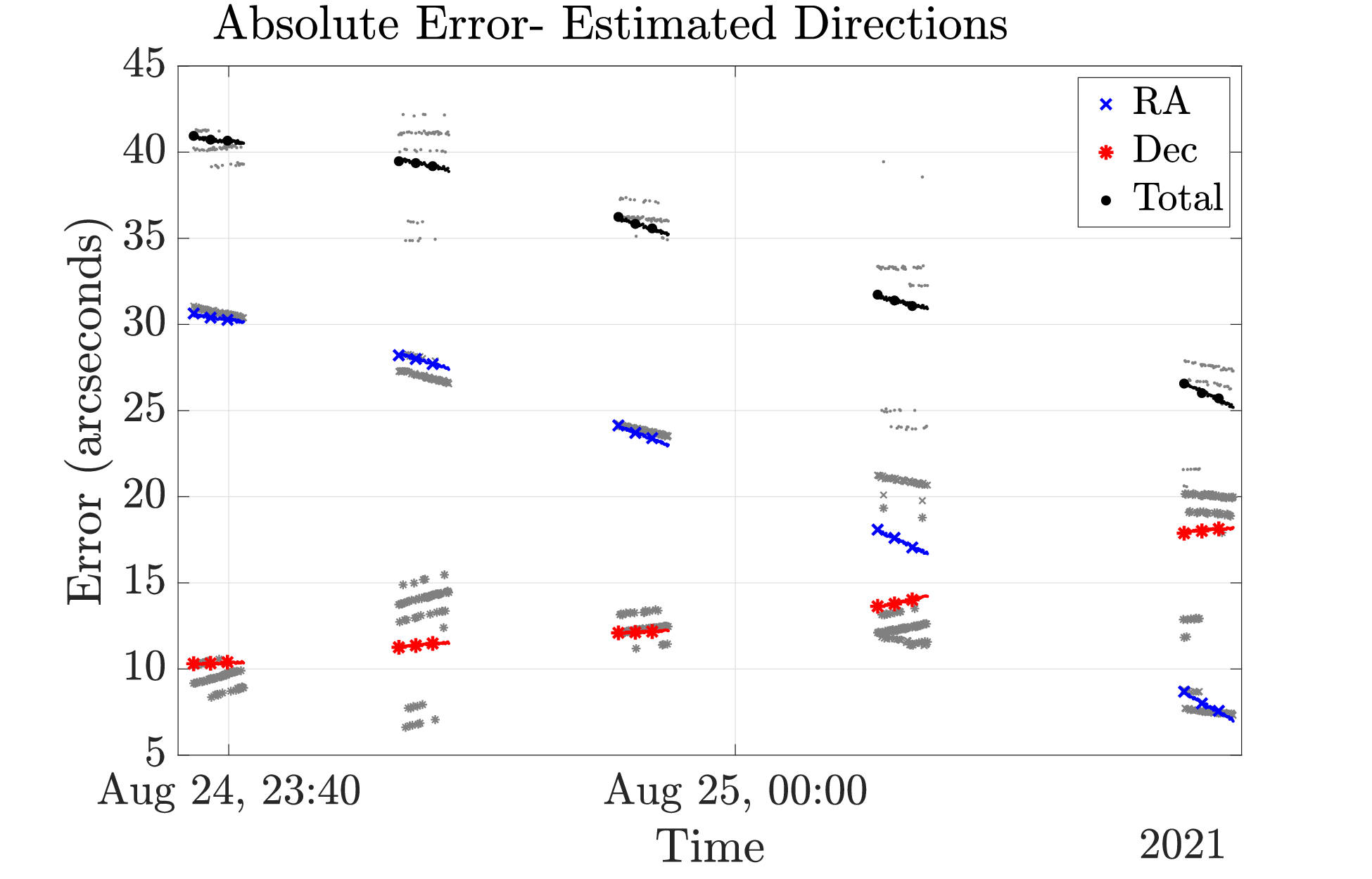}
        \caption{Absolute error of the estimated RA and Dec for the August observation.}
        \label{fig:error_estimated_dir}
    \end{figure}

\begin{figure}[!h]
\centering
        \includegraphics[width=8.5cm]{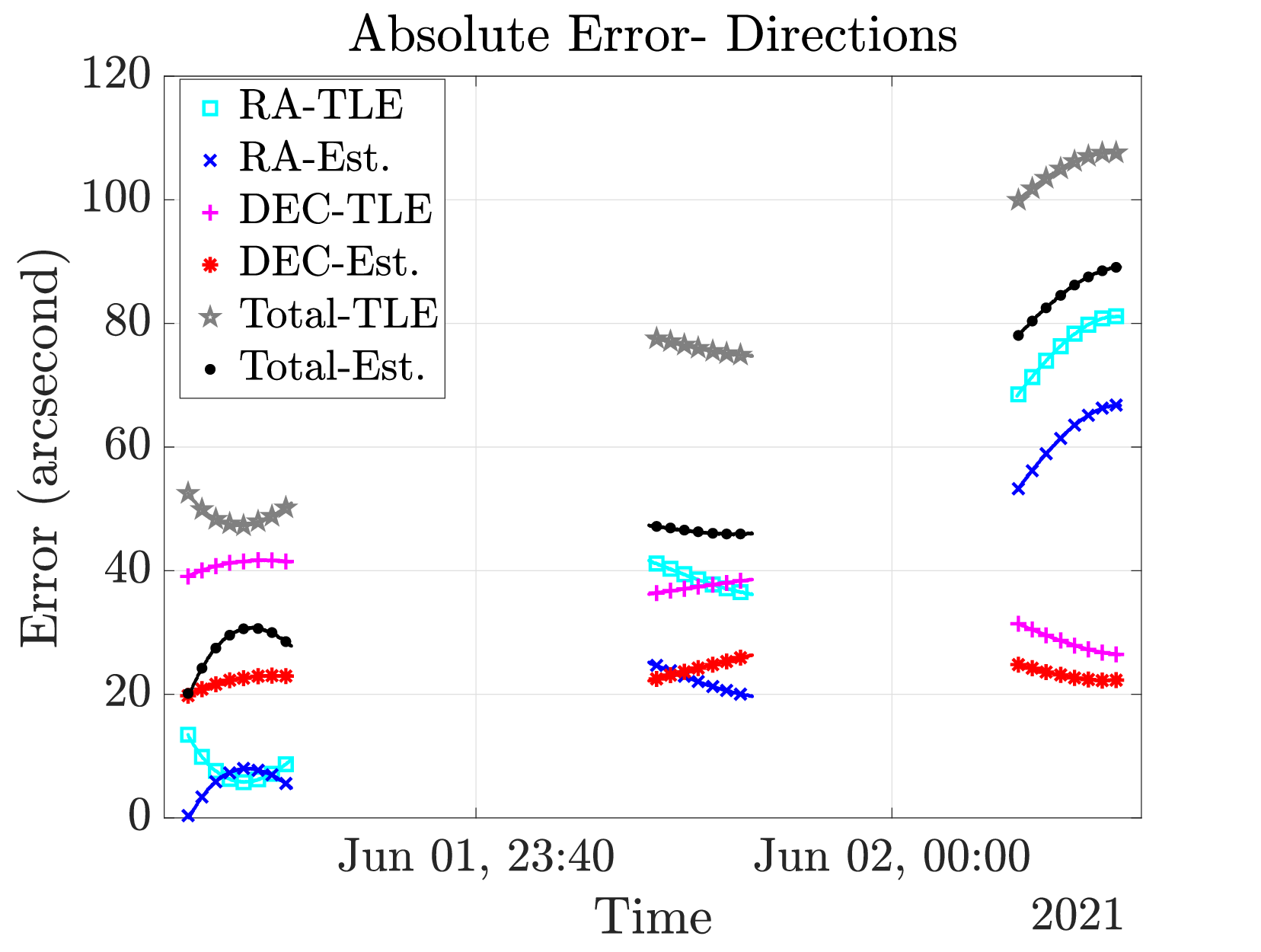} 
        \caption{Absolute error of the estimated RA and Dec and propagated orbit for June observation.}
        \label{fig:error_dir_jun}
    \end{figure}
    
        \begin{figure}[!h]
        \centering
        \includegraphics[width=8.3cm]{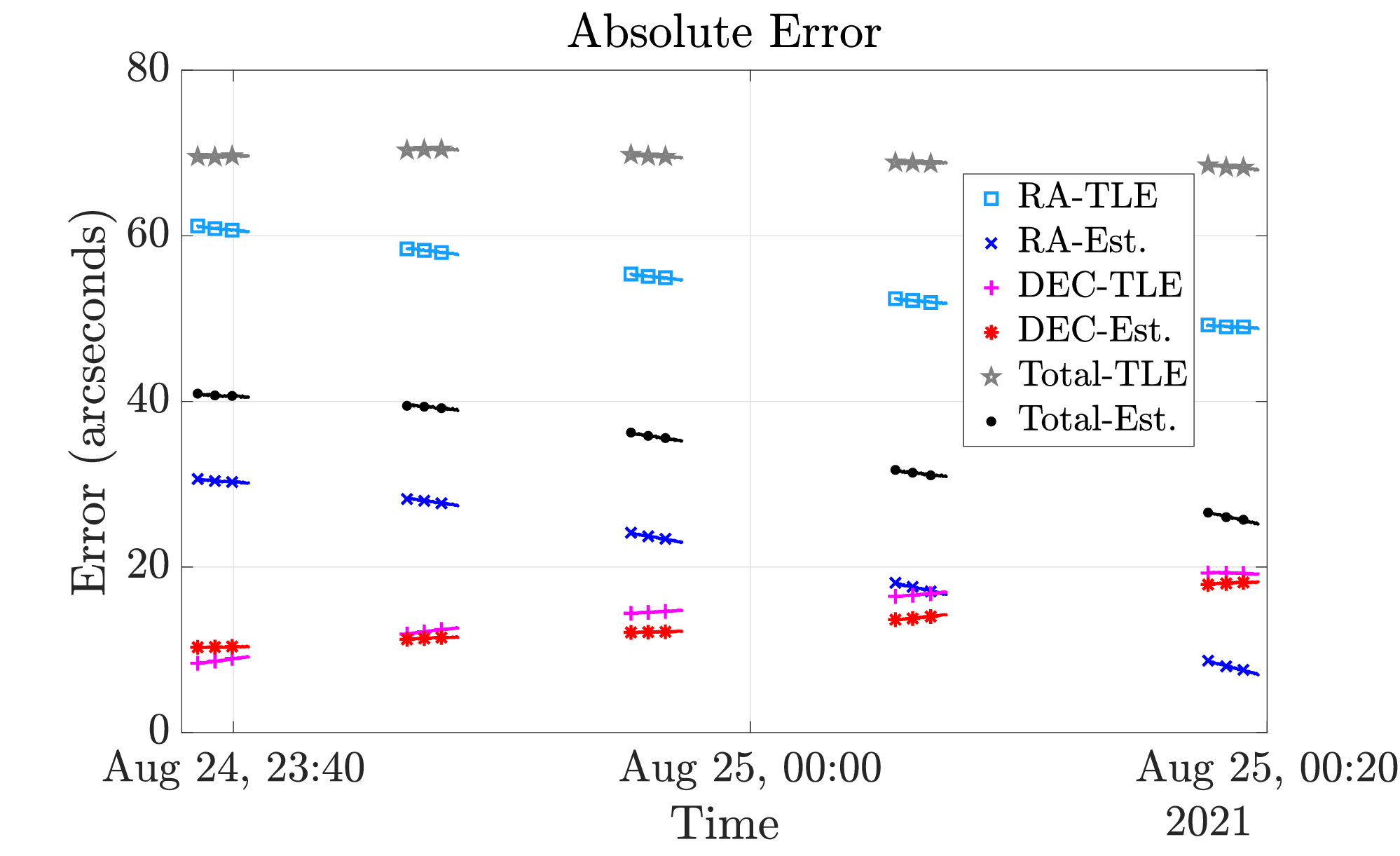} 
        \caption{Absolute error of the estimated RA and Dec and propagated orbit for the August observation.}
        \label{fig:error_dir}
    \end{figure}

Having the 3D spectrum, we locate the voxel that contains the maximum power and use the corresponding $l,m$, and $d$ coordinates as the estimated directions and range and refine the location. Then, the estimated location is compared with the location from the precise orbit. 
We interpolate the five-minute interval precise coordinate data with a second order polynomial to match our integration cycle of two seconds.
The absolute error of the estimated RA and Dec is shown in Figs.~\ref{fig:error_estimated_dir_jun} and \ref{fig:error_estimated_dir}. While there is a trend for estimated directions, the estimates are slightly different at each cycle mainly due to a slightly different signal-to-noise ratio (SNR) at each cycle, ambiguous peaks, and the discretizing effect of the applied grid. For better visibility, we fit a second-order polynomial to show the trend in color and with the actual estimates in the background. As can be seen in Figs.~\ref{fig:error_estimated_dir_jun} and \ref{fig:error_estimated_dir}, directions, i.e., RA and Dec, can be estimated with a maximum total error of approximately 90 and 41 arcseconds in June and August observations, respectively. The difference in estimation accuracy can be associated with different configurations in each experiment. The accuracy of direction estimates depends on the compact part of the array, which is approximately 75 meters in the June experiment and 214 meters in the August experiment, providing higher resolution in the latter configuration. The estimation accuracy is demonstrated to be about twice as good as the  a priori TLE data as shown in Figs.~\ref{fig:error_estimated_dir_jun} to \ref{fig:error_dir}. The mean and standard deviation of smoothed direction estimates and the TLE data are provided in Table~\ref{tab:dir_stats}, summarizing the average improvement achieved by each measurement.  


\begin{table}[!t]
\centering
    \caption{Direction absolute error statistics }
    \label{tab:dir_stats}
    \centering
    \begin{tabular}{|m{3cm}|m{1.5cm}|m{1.5cm}|}
             \hline
        Observation & Mean (arcsec)& Standard Deviation (arcsec) \\
         \hline
 Jun Est. RA  & 29.9 & 23.6 \\
   \hline
 Jun TLE RA  & 41 & 28.3 \\
\hline
 Jun Est. Dec  & 23.2 & 1.3 \\
   \hline
 Jun TLE Dec  & 35.6 & 5.4 \\
         \hline
 Aug Est. RA  & 21.4 & 8.1 \\
   \hline
 Aug TLE RA  & 55 & 4.2 \\
\hline
 Aug Est. Dec  & 13.1 & 2.7 \\
   \hline
 Aug TLE Dec  & 14.3 & 3.6 \\
\hline
    \end{tabular}

\end{table}

\begin{figure}[!h]
        \includegraphics[width=8.5cm]{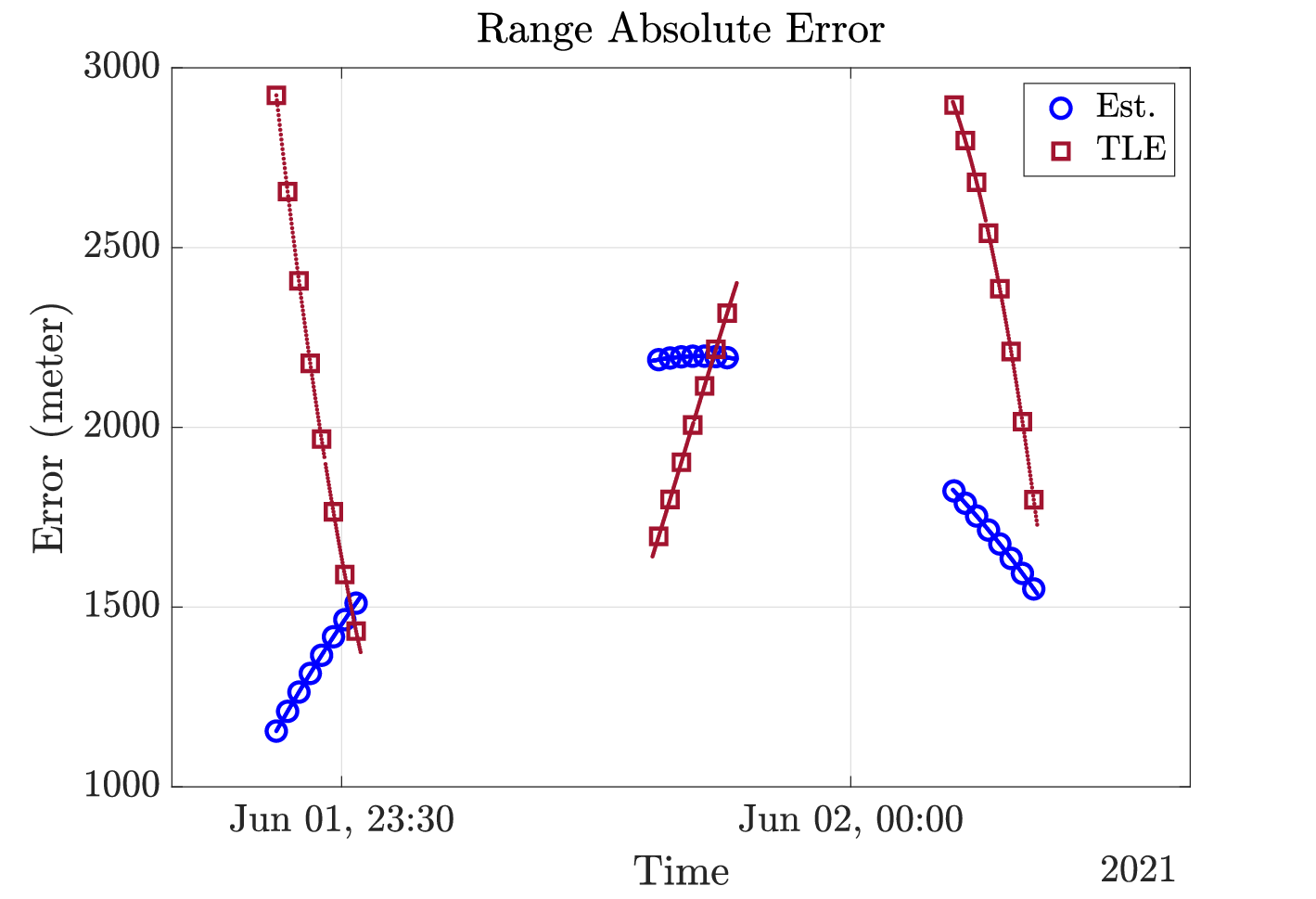} 
        \caption{Absolute error of the estimated range and propagated orbit for June observation.}
        \label{fig:range_error_jun}
            
        \includegraphics[width=8.5cm]{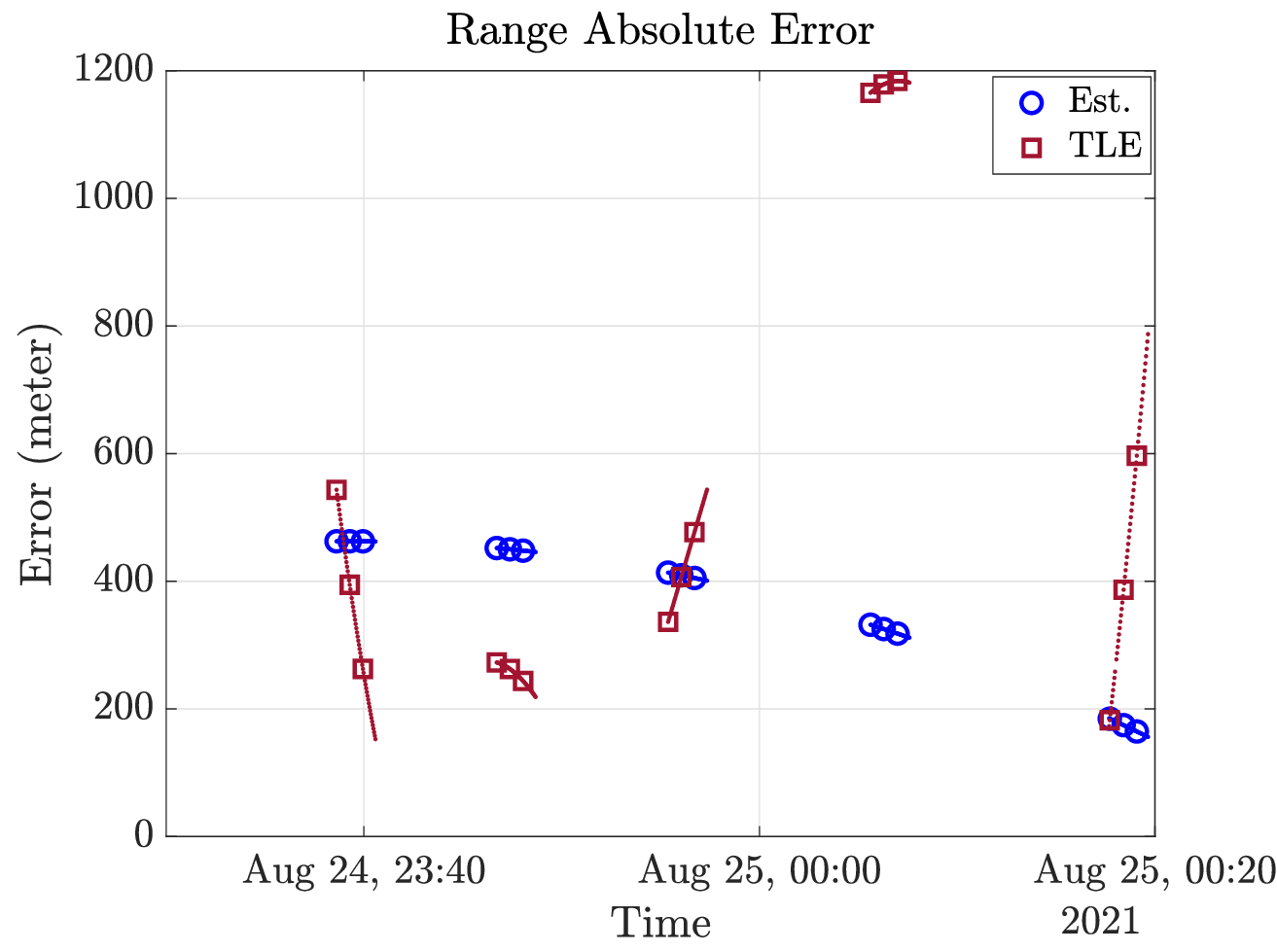} 
        \caption{Absolute error of the estimated range and propagated orbit for August observation.}
        \label{fig:range_error}

    \end{figure}

The estimated ranges are compared to those of the TLE in Figs.~\ref{fig:range_error_jun} and \ref{fig:range_error}. As with the directions, we have fitted a second-order polynomial to the estimated range values. As can be seen, the range has been estimated with an accuracy of approximately 1100 meters and 180 meters for the June and August experiments, respectively. Although estimated ranges are not always better than the propagated TLE, statistically, they improve the accuracy in mean and standard deviation senses, as listed in Table~\ref{tab:rng_stats}. 
    
\begin{table}[!t]
    \caption{Range absolute error statistics}
    \label{tab:rng_stats}
    \centering
    \begin{tabular}{|m{1.6cm}|m{1.5cm}|m{2.9cm}|}
             \hline
        Observation & Mean (m)& Standard Deviation (m) \\
         \hline
 Jun Est.  & 1744 & 356 \\
   \hline
 Jun TLE  & 2166 & 392 \\
\hline
 Aug Est.  & 363 & 107  \\
 \hline
 Aug TLE  & 533 & 345  \\
\hline
    \end{tabular}

\end{table}
\section{Conclusion}
We developed a model to estimate the range and direction of arrival (DOA) of orbital radio sources based on the interferometric data provided by a correlator radio telescope. We studied the performance of the proposed system by analyzing the data from two observation campaigns of Global Positioning System (GPS) satellites by the Australia telescope compact array (ATCA) in June and August 2021. We demonstrated that direction and range estimates are significantly more accurate than the ones provided by the orbit propagated from the most recent two-line element (TLE) compared to the location provided by the precise orbital elements.

The structure of our errors indicates that searching for and eliminating systematic errors is likely to be profitable. We are also investigating higher frequencies which bring higher resolution but raise other challenges.



\bibliographystyle{IEEEtran}
\bibliography{ref.bib}

\end{document}